\newcommand{\jpsi}{\ensuremath{J/\psi} }

\newcommand{\pT}{\ensuremath{P_T} }

\documentclass[a4paper,9.5pt]{article}
\usepackage{pos}
\usepackage{caption}
\usepackage{float} %
\usepackage[caption=false]{subfig} %
\usepackage{multicol}
\usepackage{wrapfig}

\title{Inclusive photoproduction of vector quarkonium in ultra-peripheral collisions at the LHC}

\author*[a,b]{Kate Lynch}

\affiliation[a]{School of Physics, University College Dublin,
Dublin 4, Ireland}

\affiliation[b]{Universit\'e Paris-Saclay, CNRS, IJCLab, 91405 Orsay, France}

\emailAdd{kate.lynch1@ucdconnect.ie}

\abstract{We explore the prospects of extending the LHC from a hadron-hadron to a photon-hadron collider and performing studies of inclusive quarkonium photoproduction. We perform an extrapolation from HERA data to predict the yields of photoproduced $J/\psi$ mesons in proton-lead collisions. Similarly, we 
perform an extrapolation from LHC data to model the large hadronic background.
Inclusive photoproduction can be isolated from this background in the LHC detectors, namely, ALICE, ATLAS, CMS, and LHCb by imposing constraints on the hadronic activity in the main detector and in the far-forward region. This work focuses on the CMS detector.}

\FullConference{31st International Workshop on Deep Inelastic Scattering (DIS2024)\\
 8–12 April 2024\\
Grenoble, France\\}

\begin{document}
\maketitle

\vspace{-0.4cm}\section{Introduction}
\vspace{-0.25cm}Quarkonia are bound states of heavy quarks. The description of their production is factorised into the production of the heavy-quark pair and the subsequent hadronisation of this pair into a bound state. Consequently, their study provides an opportunity to explore the strong interaction in the perturbative and non-perturbative energy regimes. There are three main approaches to quarkonium production: the Colour Singlet Model (CSM)~\cite{Chang:1979nn,Berger:1980ni,Baier:1981uk}, the Colour Octet Mechanism (COM) within non-relativistic QCD (NRQCD)~\cite{Bodwin:1994jh,Cho:1995vh,Cho:1995ce}, and the Colour Evaporation Model (CEM)~\cite{Halzen:1977rs,Fritzsch:1977ay}, none of which can simultaneously describe the inclusive hadroproduction and inclusive photoproduction data~\cite{Lansberg:2019adr}. Owing to the larger colliding energies and luminosities involved, LHC data for hadroproduction (e.g.~\cite{LHCb:2021pyk,Andronic:2015wma}) are significantly more precise than those of HERA for inclusive photoproduction (e.g.~\cite{H1:1996kyo,H1:2002voc,H1:2010udv,Kramer:2001hh}). Despite this, photoproduction data offers an enhanced discriminating power with respect to hadroproduction~\cite{Lansberg:2019adr}, particularly for the free parameters of the COM. This calls for a more precise inclusive photoproduction dataset~\cite{Boer:2024ylx}.

It is natural to study inclusive photoproduction, namely when a quasi-real photon emitted by a charged particle breaks a proton to produce a system of particles, in electron-proton colliders. Hadron-hadron collision systems, on the other hand, are associated with large hadronic backgrounds rendering these measurements challenging: the ratio of inclusive photo- to hadroproduction cross sections ranges from 0.01--0.0001 in $p$Pb collisions at nucleon-nucleon centre-of-mass energies, $\sqrt{s_{NN}}$, of $8.16$~TeV~\cite{Lansberg:2024zap}. Despite this, we demonstrate that inclusive photoproduction in hadron-hadron collisions at the LHC can be accessed by imposing a set of selection requirements. Moreover, the LHC offers certain advantages over HERA: an improved statistical precision, a deeper kinematic reach, and no contamination from the decay of $b$ hadrons with more advanced detectors.

In hadron-hadron colliders, photoproduction is typically accessed by selecting a class of events called ultra-peripheral collisions (UPCs). These are interactions mediated over distances larger than the sum of the colliding radii. At such distances ($\gtrsim1$~fm), strong interactions are suppressed and electromagnetic interactions dominate. Of course, there is no direct experimental access to the impact parameter of the collision but rather it is inferred through experimentally accessible quantities, e.g., the centrality class.

Here, we demonstrate that it is possible to measure inclusively photoproduced quarkonia at the LHC. We focus the discussion on \jpsi in a proton-lead collision system. The $J/\psi$, owing to its large cross section and significant branching to muon pairs, can be cleanly measured in the detectors of the LHC. Additionally, the lead ion has a photon flux enhanced by the charge of the ion squared, $Z^2$, with respect to a proton. We further focus on the CMS detector as it shows the most promising prospects for a \jpsi measurement.
For a more detailed study and a discussion of the capabilities of the ALICE, ATLAS, and LHCb detectors we guide the reader to the recent work~\cite{Lansberg:2024zap}.

\vspace{-0.5cm}\section{Building a Monte Carlo sample}
\label{mc}
\vspace{-0.25cm}To assess the feasibility of a photoproduction measurement at the LHC we need a MC sample for both the signal and background. We generate partonic events using \texttt{HELAC-Onia} (\texttt{HO})~\cite{Shao:2012iz,Shao:2015vga}, which is a matrix-element generator at leading order (LO) accuracy for quarkonium production within the NRQCD framework. We then interface the \texttt{HO} output to \texttt{PYTHIA}~\cite{Bierlich:2022pfr,Sjostrand:2014zea} for parton-shower and hadronisation effects. Quarkonia are sensitive to higher-order QCD corrections~\cite{Lansberg:2019adr} and parton-shower effects~\cite{Cano-Coloma:1997dvl} and there is presently no event generator for quarkonium production beyond LO. Thus, to make projections we tune the resulting \pT distributions of our MC to data. 

We model the photoproduction signal independently with a colour singlet ($^3S_1^{[1]}$)\footnote{We make use of the spectroscopic notation: $^{2S+1}L^{[c_f]}_J$, where $S$ is the spin, $L$ the angular momentum, $J$ the total angular momentum, and $[c_f]$ the colour state (1=singlet and 8=octet) of the quark--anti-quark pair.} and colour octet ($^1S_0^{[8]}$) contributions. We determine a tune by comparing $^3S_1^{[1]}$ and $^1S_0^{[8]}$ \texttt{HO+PYTHIA} results to H1 data~\cite{H1:1996kyo,H1:2002voc,H1:2010udv}, with the $b$-feed-down contribution removed by subtraction~\cite{Flore:2020jau}. This tune is then applied to \texttt{HO+PYTHIA} MC generated in proton-lead collisions at $\sqrt{s_{NN}}=$8.16~TeV.   

We model the hadronic background independently with $^3S_1^{[1]}$ and $^3S_1^{[8]}$ states\footnote{Note that we make use of a different colour octet contribution than we did for photoproduction. This is because, as demonstrated in LO studies of hadroproduction~\cite{Cho:1995vh}, the $^3S_1^{[8]}$ is the dominant octet contribution.}. Tune factors are determined by comparing the \texttt{HO+PYTHIA} results to LHCb $pp$ data at $\sqrt{s}=5$~TeV \cite{LHCb:2021pyk}. The tune is then scaled by $A$, the mass number of the lead ion, and applied to \texttt{HO+PYTHIA} MC generated in proton-proton collisions at $\sqrt{s_{NN}}=$8.16~TeV. The number of nucleon-nucleon interactions per collision, $N_\text{coll}$, differs between $pp$ and $p$Pb collisions: $N^{pp}_\text{coll}=1$ and $N^{p\text{Pb}}_\text{coll}\ge1$. To account for this, we generate minimum bias events with \texttt{PYTHIA} and fold these with the background MC according to an $N_\text{coll}$ distribution extracted from ALICE data~\cite{ALICE:2015kgk}.

\vspace{-0.5cm}\section{Photoproduction selection requirements}
\label{selection}
\vspace{-0.25cm}Measurements of photon-induced processes at the LHC have largely focused on \textit{exclusive} interactions, i.e., final states that are, in principle, fully determined. The event selection for such measurements is based on a combination of characterisation criteria: (1) \textbf{central exclusive}, i.e, the detection of the particle of interest with a veto on additional activity, (2) \textbf{colourless exchange}, and (3) \textbf{intact photon emitter}, which may be determined in a variety of ways.

For the selection of \textit{inclusive} photoproduction, criterion 1 is too strict and must be relaxed. In this section, we explore the impact on our MC of selecting events in a restricted CC (a loose implementation of criterion 3) and based on rapidity gaps (criterion 2). Additionally, we discuss how this selection can be improved at the CMS experiment by placing tighter requirements on activity in the far-forward region (criterion 3). In this region, we do not rely on the accuracy of our MC, and thus restrict ourselves to qualitative statements, which are also applicable to ALICE and ATLAS. 

Let us start with centrality. The photoproduction contribution in a sample of events can be enhanced by restricting to the most peripheral centrality class (CC). We estimate that 20\% of the most peripheral events (the 80--100\% CC) contains 6\% of hadroproduced \jpsi events~\cite{ALICE:2015kgk}. Hence the selection of this CC (which has been measured for \jpsi in proton-lead collisions~\cite{ALICE:2015kgk}) is expected to remove $94\%$ of the hadroproduction background and contain 100\% of the photoproduction signal.

\begin{wrapfigure}{r}{0.4\textwidth}
        \centering
        \includegraphics[trim= 0cm 0cm 0cm 0cm,clip, width=0.4\textwidth]{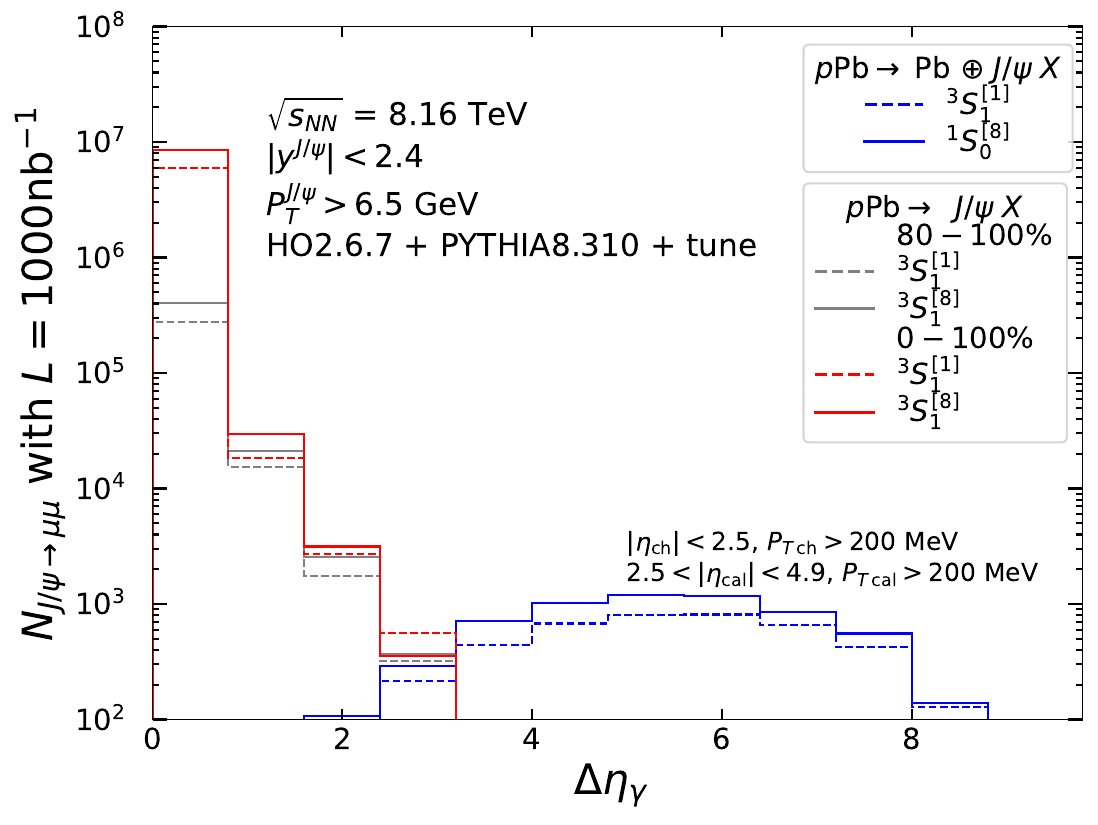}
     
         \vspace{-0.25cm}\caption{Yield for $J/\psi\rightarrow \mu \mu$ as a function of $\Delta \eta_\gamma$ in the CMS acceptance using projected Run3+4 luminosity, for the singlet (dashed) and octet (solid) tunes of photoproduction (blue) and hadroproduction in the 80--100\% (grey) and 0--100\% (red) CC. }
 \label{fig:CMS-rapgap}
\end{wrapfigure}

As for rapidity gaps, we define $\Delta \eta_\gamma \equiv \min\{|\eta_{\gamma\text{-edge}}-\eta_i |\}$ for $i\neq \jpsi$, where $\eta_{\gamma\text{-edge}}$ is the pseudorapidity of the edge of the detector on the lead-going side and $\eta_i$ are the psuedorapidities of particles within the detector acceptance, which for CMS corresponds to a charged track acceptance of $\eta_\text{cal}<2.5$, $p_{T\text{ cal}}>0.4$~GeV and a calorimeter acceptance of $\eta_\text{cal}<2.5$, $p_{T\text{ cal}}>0.4$~GeV. Figure~\ref{fig:CMS-rapgap} shows the $J/\psi$ yield as a function of $\Delta \eta_\gamma$ within the CMS acceptance in proton-lead collisions for the singlet  (dashed) and  octet  (solid) tunes of photoproduction (blue) and hadroproduction in the 80--100\% (0--100\%) CC (grey (red)). More central (equivalently larger $N_\text{coll}$) background events contain more hadronic activity and thus are easier to separate from photoproduction. This can be seen in Fig.~\ref{fig:CMS-rapgap}, where a veto on the 0--80\% CC only affects the normalisation of the first bin in $\Delta \eta_\gamma$ (compare red and grey histograms). We propose selecting events with $\Delta \eta_\gamma>\Delta \eta_\gamma^\text{stat min}$, where $\Delta \eta_\gamma^\text{stat min}$ minimises the statistical uncertainty of the sample in the selection region. The same applies for ATLAS.

The zero degree calorimeters (ZDCs) are installed on both sides of the interaction point and are sensitive to neutral particles with $\eta \gtrsim 8$. The main source of these neutral particles is neutrons coming from Pb ions that are broken during hadronic collisions. In fact, these neutrons are used to classify the centrality of a collision, more neutrons correspond to a more central collision. However, the neutrons passing through the ZDCs can also come from the photonuclear excitation and de-exciation of an ion. Following methods described in \cite{Baltz:2002pp,Broz:2019kpl}, we estimate that the probability for one or more of such neutron emissions in a photoproduced $J/\psi$ event in $p$Pb collisions is $\mathcal{O}(0.01\%)$. Thus, vetoing neutron emissions (as was done in previous UPC studies~\cite{ATLAS:2021jhn,ATLAS:2024mvt}) has an extremely high efficiency for retaining signal.

\begin{figure}[!hbt]
    \centering
     \subfloat[]{\label{cmsA}\includegraphics[width=0.3\textwidth]{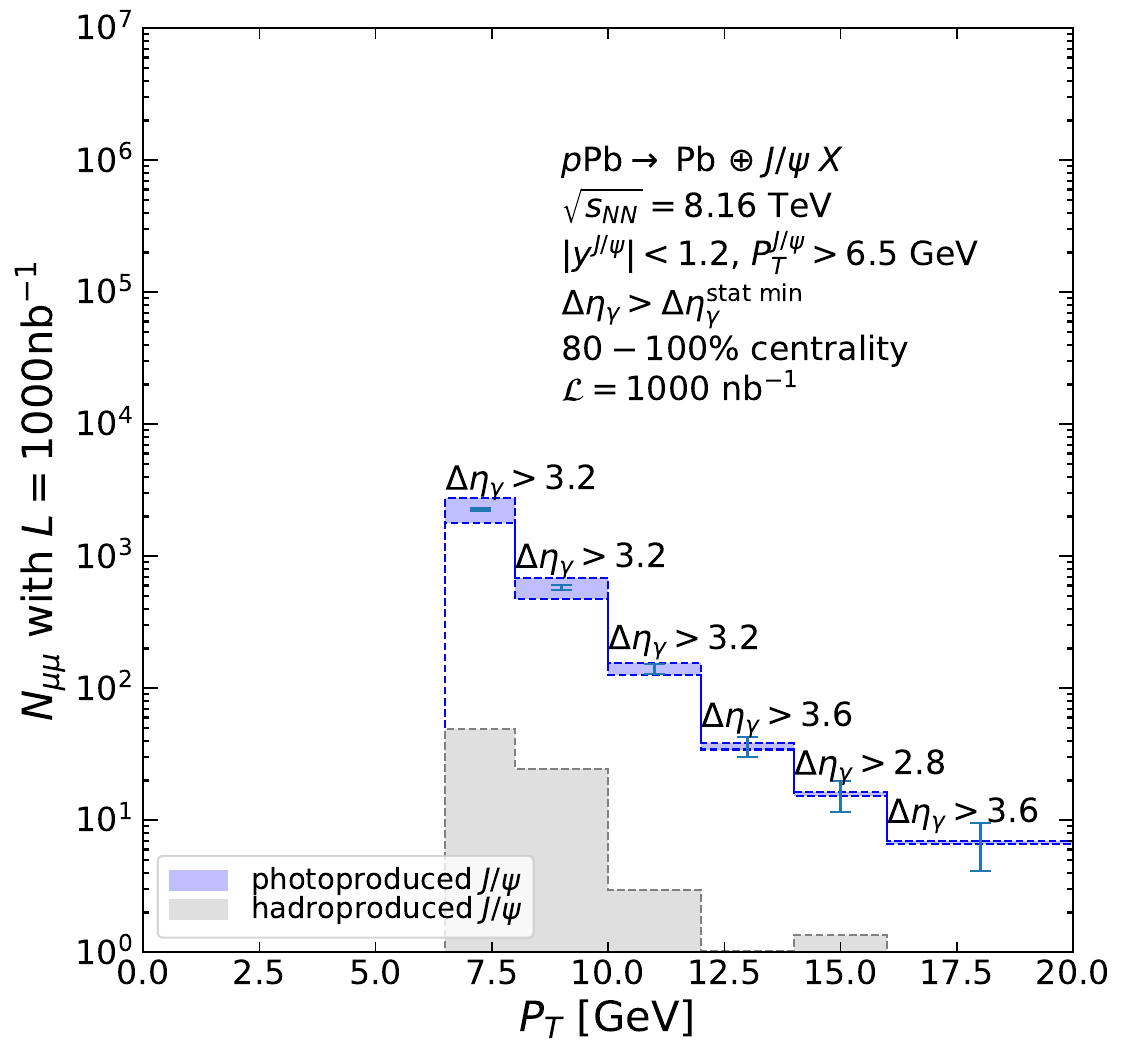}}
     \subfloat[]{\includegraphics[width=0.3\textwidth]{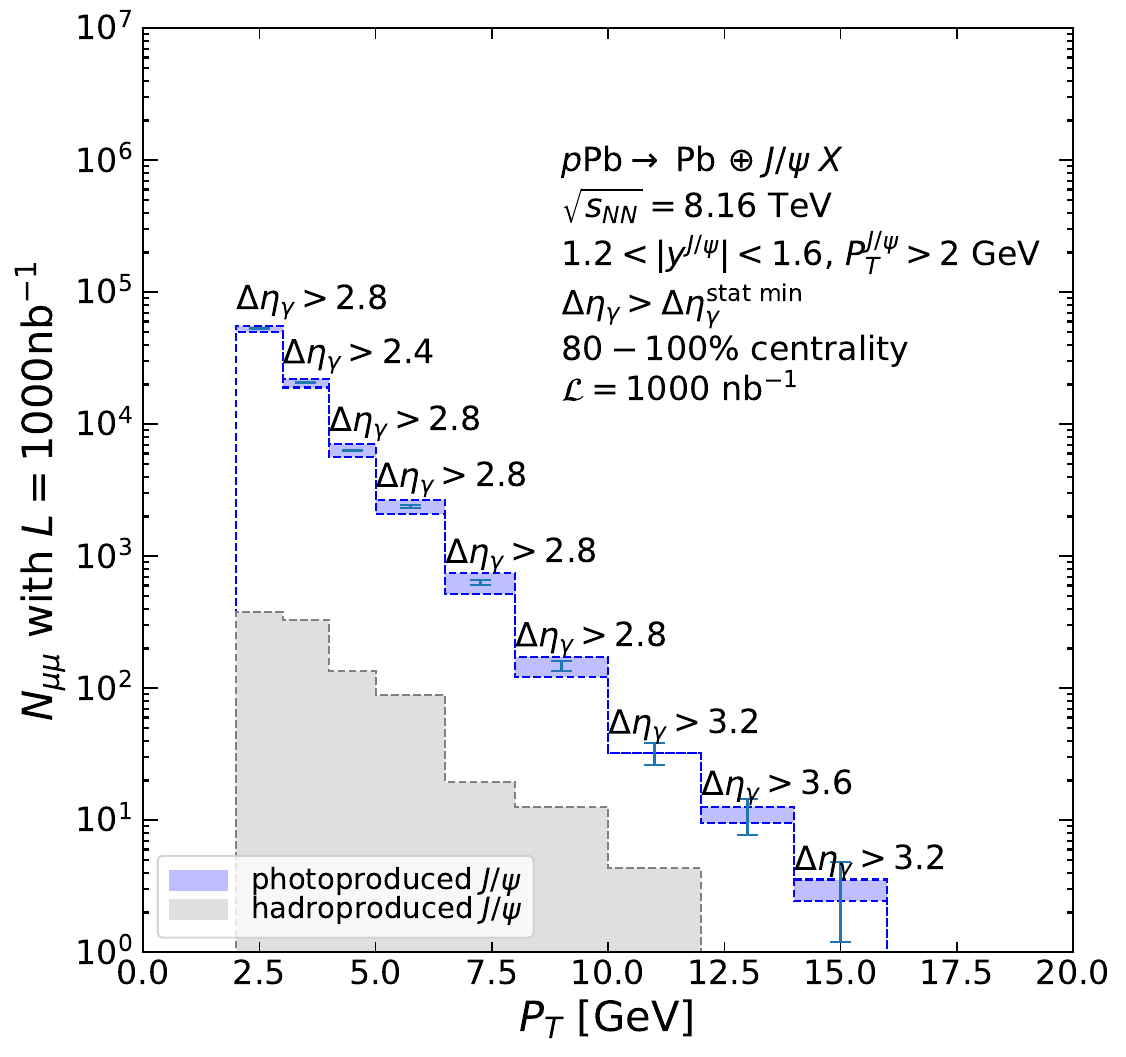}}
     \subfloat[]{\includegraphics[width=0.3\textwidth]{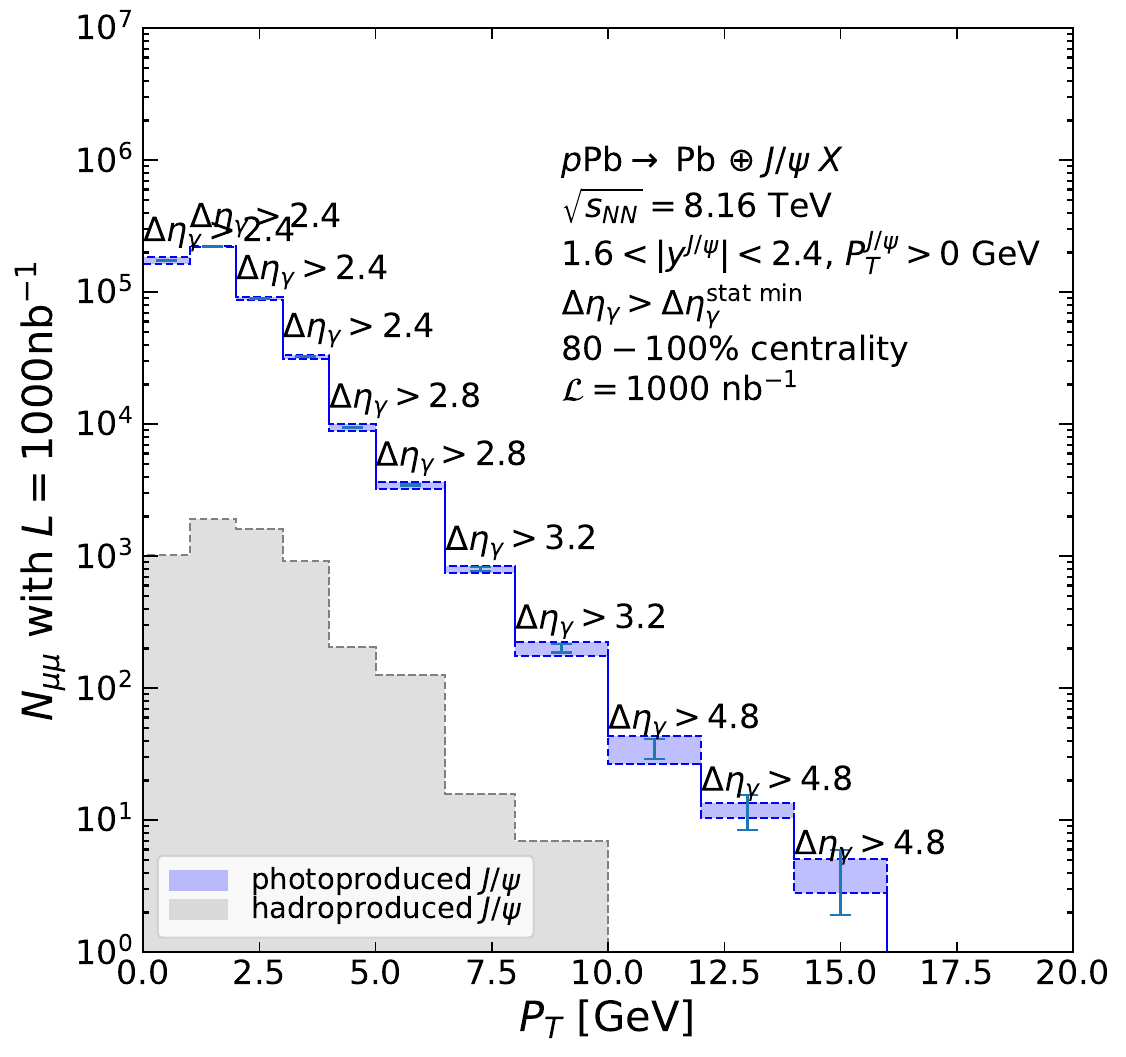}}
   \vspace{-0.15cm}\caption{The $J/\psi$ yield and associated statistical uncertainty (vertical error bars) using projected Run3+4 luminosity after selection requirements as a function of \pT for photoproduction (blue) and hadroproduction (grey) with \jpsi in the CMS acceptance with (a) $|y^{\jpsi}|<1.2$, (b) $1.2<|y^{\jpsi}|<1.6$, and (c) $1.6<|y^{\jpsi}|<2.4$. The coloured bands indicate the tune uncertainties. }
    \label{fig:ptaftercuts}
\end{figure}

Figure~\ref{fig:ptaftercuts} shows the predicted yield of photoproduced \jpsi (blue) and the hadroproduced background (grey) after selecting the 80--100\% CC and requiring $\Delta \eta_\gamma>\Delta \eta_\gamma^\text{stat min}$. The slope of the photoproduction signal falls more steeply in \pT than that of the hadroproduction background and so we compute $\Delta \eta_\gamma^\text{stat min}$ per bin in \pT (values are indicated on the figures). Yields are given for \jpsi in the CMS acceptance in three rapidity regions: (a) $|y^{\jpsi}|<1.2$, (b) $1.2<|y^{\jpsi}|<1.6$, and (c) $1.6<|y^{\jpsi}|<2.4$, each associated with a different \pT acceptance. As can be seen in Fig~\ref{cmsA}, there are $\mathcal{O}(10)$ events for $16<\pT<20$~GeV. A similar yield is expected for the ATLAS detector in rapidity region (a).
\vspace{-0.5cm}

\vspace{-1cm}\section{Conclusion and outlook}\label{conclusion}
\vspace{-0.25cm}We have demonstrated that hadron beams can be used as photon beams for the study of inclusive processes. 
The focus of this study was on \jpsi photoproduction in $p$Pb collisions, however, the techniques discussed here may be generalised to many processes. More precise and further reaching inclusive quarkonium photoproduction data will provide the opportunity to better constrain the quarkonium-production mechanism.

We have illustrated that the inclusive photoproduction signal can be isolated with respect to the large competing hadroproduction background in the CMS detector by imposing selection criteria based on centrality-classes and rapidity gaps. The same selection can be employed for the ATLAS selection. Using Run3+4 data, the maximum measurable \pT is expected to be 20~GeV (twice that of the existing HERA dataset). As discussed, placing a no-neutron requirement is expected to further enhance the signal purity.

\vspace{-0.5cm}\acknowledgments
\vspace{-0.3cm}I would like to thank J.P. Lansberg, R. McNulty, and C. Van Hulse for their collaboration in this work. The research conducted in this publication was funded by the Irish Research Council under grant number GOIPG/2022/478. This project is supported by the European Union's Horizon 2020 research and innovation program under Grant agreement no. 824093.
This research has also received funding through the Joint PhD Programme of Universit\'e Paris-Saclay (ADI).

\bibliographystyle{utphys}

\vspace{-0.45cm}{\footnotesize
\bibliography{bib}}
 
\end{document}